\documentclass[a4paper,twoside]{article}
\newcommand{\affil}[1]{$^{\rm #1}$}
\usepackage[authoryear]{natbib}
\bibpunct{(}{)}{;}{a}{}{,}
\usepackage{graphicx}
\usepackage{rotating}
\date{} 
\title{\large\bf\flushleft Barium Stars - Theoretical Interpretation}
\author{\parbox{\textwidth}{\flushleft
\vspace{-0.5cm}
{\it Laura Husti\affil{A,B,E}, Roberto Gallino\affil{A}, Sara Bisterzo\affil{A}, Oscar Straniero\affil{D}, and 
Sergio Cristallo\affil{D}}\\
\vspace{0.4cm}
{\small \affil{A}\,Dipartimento di Fisica Generale, Universit\`{a} degli Studi di Torino, \\ via P. Giuria 1, 10125 Torino, Italia}\\
{\small \affil{B}\,Research Centre for Atomic Physics and Astrophysics, University of Bucharest,\\
P.O.Box MG-6, RO-077125 Bucharest-Magurele, Romania}\\
{\small \affil{D}\,INAF Osservatorio Astronomico di Collurania, via M. Maggini, 64100 Teramo,
Italy}\\
{\small \affil{E}\,Email: lotesileanu@yahoo.com}}}
\begin{document}
\twocolumn[
\begin{changemargin}{.8cm}{.5cm}
\begin{minipage}{.9\textwidth}
\vspace{-1cm}
\maketitle
%
%
\small{\bf Abstract:}
Barium stars are extrinsic Asymptotic Giant Branch (AGB) stars. 
They present the s-enhancement characteristic for AGB and post-AGB 
stars, but are in an earlier evolutionary stage (main sequence dwarfs, 
subgiants, red giants). They are believed to form in binary systems, 
where a more massive companion, evolved faster, produced the 
s-elements during its AGB phase, polluted the present barium star 
through stellar winds and became a white dwarf. 
The samples of barium stars of Allen \& Barbuy (2006)
and of Smiljanic et al. (2007) are analysed here.
Spectra of both samples were obtained at high-resolution
and high S/N.
We compare these observations with AGB nucleosynthesis models
using different initial masses and a spread of $^{13}$C-pocket 
efficiencies. 
Once a consistent solution is found for the whole elemental
distribution of abundances, a proper dilution factor is applied. 
This dilution is explained by the fact that the $s-$rich material 
transferred from the AGB to the nowadays observed star, is mixed 
with the envelope of the accretor. We also analyse the 
mass transfer process, and obtain the wind velocity for giants and 
subgiants with known orbital period. We find evidence that 
thermohaline mixing is acting inside main sequence dwarfs and we 
present a method for estimating its depth. 


\medskip
\medskip
\end{minipage}
\end{changemargin}
]
\small

\section{Introduction}

Barium stars have been a fascinating subject since their discovery 
by Bidelman \& Keenan (1951). Understanding their peculiar abundance 
patterns has been a challenge for over 50 years. 

Barium stars show s-process enhancement.
However, these stars are giants or main-sequence, far away from
the thermal pulse asymptotic giant branch stars (TP-AGB) phase,
where the s-process is manufactured. 
Late on the TP phase, AGB stars undergo recurrent third dredge-up
(TDU) episodes, mixing with the envelope C-rich and s-process-rich 
material synthesised in the He intershell (the zone between
H shell and He shell).
The most likely interpretation of barium stars is that they acquired
the s-enrichment by mass transfer in a binary system
through stellar winds from an AGB companion that became later 
on a white dwarf (in some cases detectable, see Boffin \& Jorissen 
1988, Jorissen \& Mayor 1988 and references therein). 
The study of barium stars is important for a better understanding 
of the s-process in AGB stars, of the mass transfer in binary 
systems and of the mixing processes, like thermohaline mixing, 
which may occur in the envelope of main sequence stars
(Stancliffe et al. 2007 and references therein). 
This is a process that occurs when the mean 
molecular weight of the
stellar material increases towards the surface. The heavier 
material transferred from the
AGB will sink until the difference in molecular  
gradient weight will disappear. 

A recent homogeneous sample of 26 barium stars with lead measurement
detected at high-resolution spectroscopy (S/N $\sim$ 100 to 250)
 has been analysed by Allen \& Barbuy (2006).
 About half of these stars are barium dwarfs, and half are barium
 subgiants and giants.
Many heavy elements have been observed, among which, Sr, Y, Zr,
belonging to the first s-process peak (ls, light-s, at neutron 
magic number N = 50), Ba, La, Ce, Pr, Nd, Sm, belonging to the 
second s-peak (hs, heavy-s, at N = 82).
Moreover, for the first time, Pb at the termination 
of the s-process (N = 126), was detected. 
Similar high-quality spectra (with higher S/N $\sim$ 500) for 
other 8 barium giants has been analysed by Smiljanic et al. (2007),
without lead measurement.
The aim of this paper is to compare our theoretical predictions
with Allen \& Barbuy (2006) and Smiljanic et al. (2007)
observational samples.

In Section 2 we present the models used.
In Section 3 we compare our theoretical predictions with the 
observations and we discuss in detail the case of a barium dwarf
(HD 123585) and of a barium giant (HD 27271).
In Section 4 we perform 
mass transfer calculations for some stars with known orbital period. 
These results allow us to check the self-consistency of the
dilution factor obtained for giants (or subgiants) and to estimate 
the depth of the thermohaline mixing for barium dwarfs. 
 

\section{Theoretical models}

\begin{figure}[htbp]
  \resizebox{75mm}{!}{\includegraphics[angle=-90]{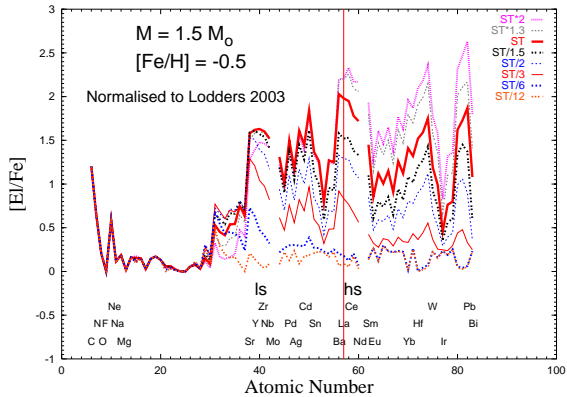}}
\caption{Theoretical predictions of [El/Fe] for AGB models
of $M$ = 1.5 M$_\odot$, [Fe/H] = $-$0.5 and various $^{13}$C-pocket 
efficiencies (from ST$\times$2 down to ST/12). We normalised to the 
photosperic values by Lodders (2003).}
 \label{hustipasa_sara1}
\end{figure}

\begin{figure}[htbp]
  \resizebox{75mm}{!}{\includegraphics[angle=-90]{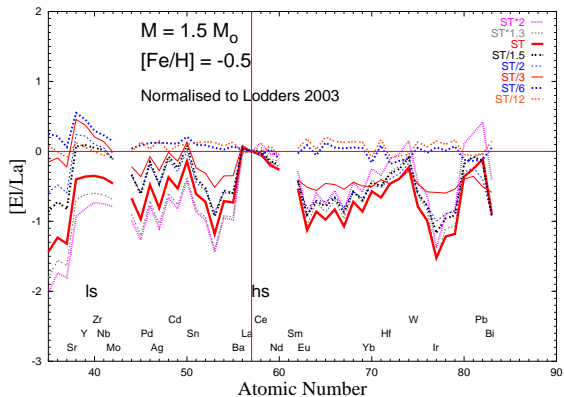}}
\caption{Theoretical predictions [El/La] for elements from Sr to 
Bi for AGB models of $M$ = 1.5 M$_\odot$, [Fe/H] = $-$0.5 and various 
$^{13}$C-pocket efficiencies (from ST$\times$2 down to ST/12).}
 \label{hustipasa_sara2}
\end{figure}

Our models are based on FRANEC
(Frascati Raphson-Newton 
Evolutionary Code, Chieffi \& Straniero 1989), coupled with a 
post-process code that includes a full network up to bismuth.
According to FRANEC, TDU starts after
a limited number of TPs. TDU ceases when the
mass of the envelope decreases by mass loss down to $\sim$ 0.5 $M_\odot$.
The TDU mass per pulse reaches a maximum after about ten 
thermal pulses and then decreases (Straniero et al. 2003). 
The major neutron source in AGB stars is the 
$^{13}$C($\alpha$, n)$^{16}$O reaction,
which releases neutrons in radiative conditions during the interpulse
phase (Straniero et al. 1995).
During TDU, a small amount of protons
is assumed to penetrate in the top layers of the He intershell.
At H reignition, protons are captured by the abundant $^{12}$C,
via $^{12}$C(p, $\gamma$)$^{13}$N($\beta^+$$\nu$)$^{13}$C.
We use AGB models having initial masses of 1.4, 1.5, 2 and 3 
M$_\odot$, different metallicities and a range of $^{13}$C-pocket 
efficiencies, from ST $\times$ 2 down to ST/12. Case ST is the 
standard choice of Gallino et al. (1998), which was shown
to best reproduce the solar main component of the s-process
(Arlandini et al. 1999). The ST $\times$ 2 case roughly corresponds 
to the maximum $^{13}$C-pocket efficiency. Higher proton abundances  
would lead to the formation of $^{14}$N at expenses of $^{13}$C. 
Moreover, $^{14}$N would act as a major neutron poison.
We may define a minimum $^{13}$C-pocket as the one that significantly 
affects the final $s$-process distribution. 
This limit corresponds roughly to ST/6 at solar metallicity and
decreases with decreasing metallicity, given that the neutron 
exposure is inversely proportional to the iron seeds.
A second minor neutron source is driven by the reaction
 $^{22}$Ne($\alpha$, n)$^{25}$Mg, 
partially activated during TPs, producing a neutron 
burst of small neutron exposure and high peak neutron density. 
For a detailed description of the models we refer to Busso, Gallino
\& Wasserburg (1999) and Straniero et al. (2003). 

As barium stars owe their s-enhancement to mass transfer, 
the s-rich material from the AGB donor was mixed and diluted
with the envelope of the future barium star. 
Defining the dilution factor as 
$M^{env}_{\star}$/$M^{transf}_{\rm AGB}$ = 10$^{dil}$, 
where $M^{env}_{\star}$ is the mass of the envelope 
of the observed star after the mass transfer,
$M^{transf}_{\rm AGB}$ is the mass transferred from the AGB,
we obtain the diluted theoretical abundances\footnote{The 
standard spectroscopic notation is adopted: [A/B] = log$_{10}$(N$_{\rm A}$/N$_{\rm B}$)$_{\star}$ - log$_{10}$(N$_{\rm A}$/N$_{\rm B}$)$_{\odot}$.}:

\begin{equation}
\left[\frac{\rm X}{\rm Fe}\right]={\rm log}\left(10^{\left[\frac{\rm X}{\rm 
 Fe}\right]^{ini}}\cdot f + 10^{\left[\frac{\rm X}{\rm Fe}\right]^{\rm AGB}} 
 \cdot 10^{-dil}\right),  
\end{equation}

where $f = 1-10^{-dil}$ and $\rm [X/Fe]^{\rm AGB}$ 
is the abundance of the element $\rm X$ in the AGB. 
We suppose that the two stars
formed from the same cloud of interstellar gas. 

Dilution will lower the abundances of the elements, but will 
not change the shape of the s-process distribution. If $dil$ $\simeq$ 
0, it means that no mixing had occurred after the mass transfer, 
and the observed abundances are the same as in the AGB star.

We present in Fig.~\ref{hustipasa_sara1} 
theoretical predictions for elements from carbon to bismuth using
AGB models of 1.5 $M_\odot$ and [Fe/H] = $-$0.5 at the last TDU episode
for a range of $^{13}$C-pocket efficiencies.
Note that about half of the AGB mass ejected by stellar winds has
a chemical composition shown in the Figure. Indeed, 
as recalled in Section~2, according to FRANEC,
TDU ceases when the envelope mass is of the order of 0.5 $M_\odot$.
The same results are shown in Fig.~\ref{hustipasa_sara2} for
heavier elements normalised to lanthanum, which is the best 
representative element of the hs peak.

\section{Results}
The Allen \& Barbuy (2006) sample contains 
both dwarfs and barium giants or subgiants. 
Barium dwarfs have negligible convective
envelopes.
In order to estimate the mass of convective envelope
of our sample stars, we used the Window To The Stars (WTTS) 
interface to Peter Eggleton's TWIN single and binary stellar 
evolution code (Izzard \& Glebbeek 2006).
Giants and subgiants have larger convective envelopes 
and some of the giants have already suffered the first 
dredge-up (FDU), so we expect large dilution factors
in these cases. 

In order to find a possible solution for a barium star, 
we seek for the $^{13}$C-pocket efficiency that would reproduce
the observed trend for the elements belonging to the three s-peaks, 
ls, hs and Pb. We define 
\begin{equation}
[{\rm ls}/{\rm Fe}] = \frac{([{\rm Y}/{\rm Fe}] + [{\rm Zr}/{\rm Fe}])}{2} 
\end{equation}
and 
\begin{equation}
[{\rm hs}/{\rm Fe}] = \frac{([{\rm La}/{\rm Fe}] + [{\rm Nd}/{\rm Fe}] + [{\rm Sm}/{\rm Fe}])}{3}
\end{equation}
and their ratios [hs/ls] and [Pb/hs], two 
s-process indexes independent of the dilution factor. 

In Fig.~3--5, we compare spectroscopic observations of 
the barium dwarf HD 123585 and the barium 
giant HD 27271 by Allen \& Barbuy 
(2006)\footnote{The errorbars in the plots 
are quoted from Allen \& Barbuy (2006) and
represent the highest value between the uncertainty due to the 
atmospheric model together with the stellar parameters and 
the line-to-line scatter for elements with more than two lines 
used to derive the abundances. These errorbars are plotted 
as solid lines. For elements where the abundances were 
derived from only one or two lines 
(O, 2 lines; 
Sr, 1 line for dwarfs, 2 lines for giants;
Pb, 1 line; Mo, 1 line) 
we have plotted with dashed line the 
errorbar. 
This value has to be considered as a lower limit of the 
uncertainty for the abundances of these elements.}. 

\begin{figure}[htbp]
  \resizebox{75mm}{!}{\includegraphics[angle=-90]{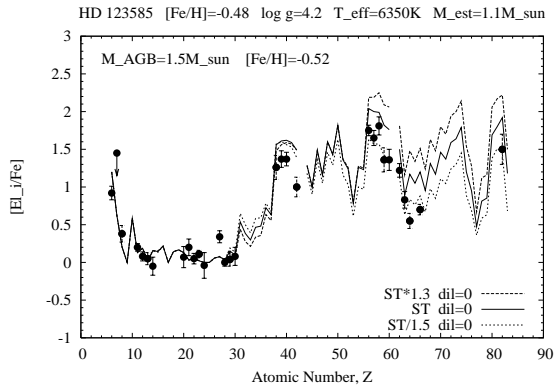}}
\caption{The barium dwarf star HD 123585 elemental spectroscopic 
distribution compared with theoretical predictions. We plot the 
theoretical abundances without dilution, for an AGB with
initial mass $M_{\rm AGB}$ = 1.5 M$_\odot$, 
 [Fe/H] = $-$0.52 and three different $^{13}$C-pocket 
efficiencies (ST, ST $\times$ 1.3 and ST/1.5).}
 \label{compnodil}
\end{figure}


\begin{figure}[htbp]
  \resizebox{75mm}{!}{\includegraphics[angle=-90]{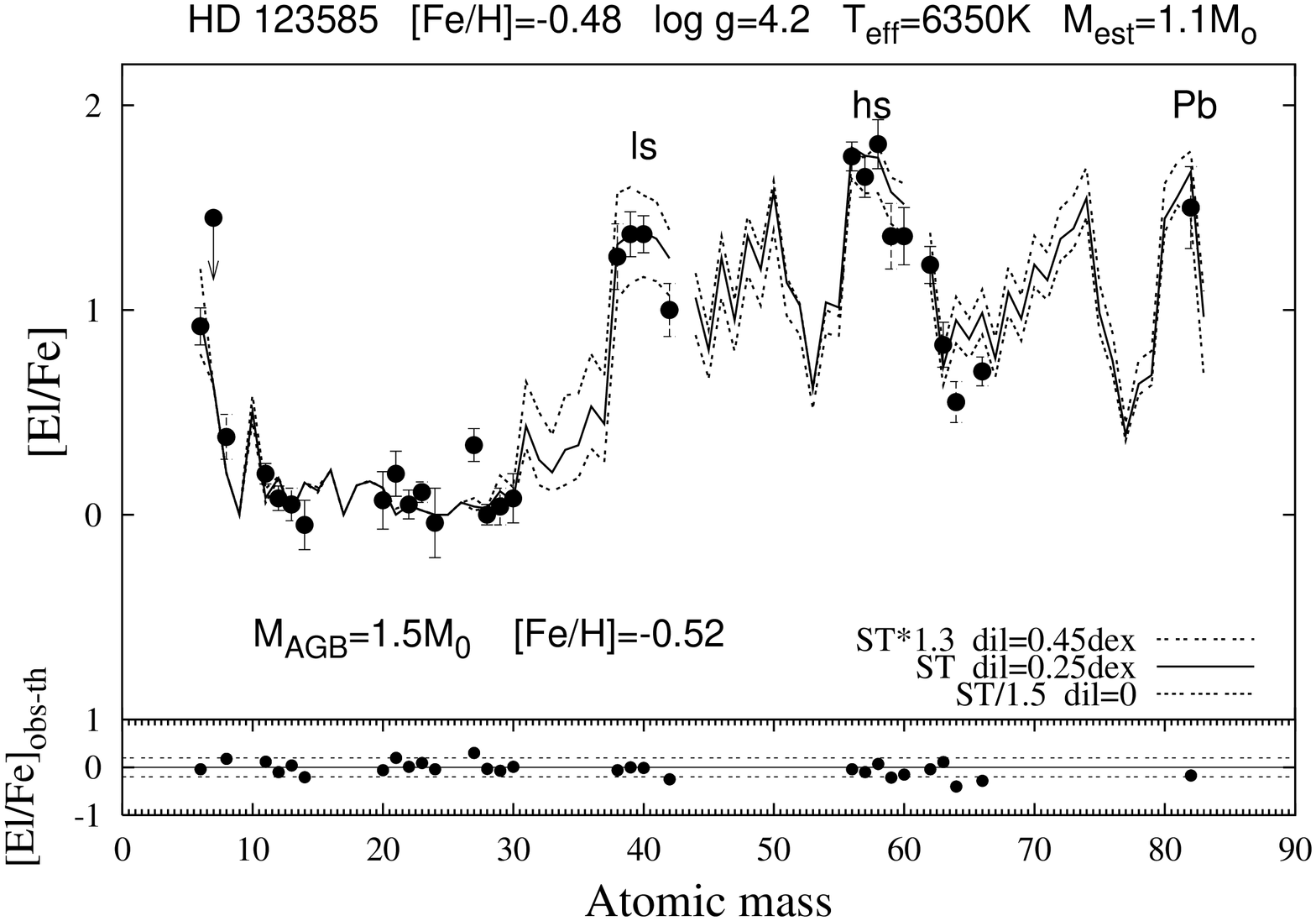}}
\caption{The barium dwarf star HD 123585 elemental spectroscopic 
distribution compared with theoretical predictions. We plot the 
theoretical abundances with dilution, for the 1.5-M$_\odot$ 
AGB model with [Fe/H] = $-$0.52 and three $^{13}$C-pocket 
efficiencies (ST, ST $\times$ 1.3 and ST/1.5).}
 \label{compdil}
\end{figure}


\textbf{HD 123585} is a \textbf{barium dwarf} of metallicity 
[Fe/H] = $-$0.48 with orbital period of 458 days (Pourbaix 
\& Jorissen 2000).
The s-enhancement is quite 
large: [ls/Fe] $\sim$ 1.4 dex, [hs/Fe] $\sim$ 1.7 
dex and [Pb/Fe] $\sim$ 1.5 dex.
For HD 123585, the ST case (the solid line in 
Fig.~\ref{compnodil}) satisfies this condition: in the 
theoretical case [hs/ls]=0.14 and [Pb/hs]=0.17, while from 
the observations we have [hs/ls] and [Pb/hs] respectively 0.07 
and 0.13, with a typical uncertainty of 0.2 dex. For the other 
two cases plotted in Fig.~\ref{compnodil}, ST $\times$ 1.3 and 
ST/1.5, the discrepancy in [hs/ls], with respect to the 
observational values is $\sim$ 0.4 dex for the ST $\times$ 
1.3 case, and $\sim$ 0.3 dex for the ST/1.5 case. 
The dilution factor $dil$ is then chosen in 
order to match the [hs/Fe] predictions with spectroscopic data.
%
In Fig.~\ref{compdil} we notice that, after dilution, all the 
three cases shown (ST$\times$1.3, ST and ST/1.5) match the 
hs-peak, but only the ST case matches the ls-peak.

%

Very similar solutions are obtained with the ST case and an 
initial mass of the AGB donor of 2 and 3 M$_\odot$.
The dilution factor is 0.25 dex for the 1.5-M$_\odot$ model, 0.6 
dex for the 2-M$_\odot$ model and 0.5 for the 3-M$_\odot$ model.  
Mo, Gd and Dy, each derived from 1 line, are lower than expected. 
Note that our relative s-process expectations of close-by elements
are quite robust and are based on accurate neutron capture cross sections.

\textbf{HD 27271} is a \textbf{barium giant} of solar metallicity
with orbital period of 1694 days (Udry et al. 1998a). 
For this star [ls/Fe] $\sim$ 1, [hs/Fe] $\sim$ 0.5 and
[Pb/Fe] $\sim$ 0.3 dex. We were able to reproduce the observations
using a case ST $\times$ 1.3,
and dilution factors 0.1, 0.3 and 
0.45 dex for models with $M$ = 1.5, 2 and 3 M$_\odot$, respectively.  
The solution found for the 1.5-M$_\odot$ initial mass of 
the AGB donor was discarded, because the estimated mass 
of this star is M$_{\mathrm{est}}$ = 1.9 M$_\odot$ (Allen
\& Barbuy 2006). 
Mo, with only one line detected, appears too low.



\begin{figure}[htbp]
  \resizebox{75mm}{!}{\includegraphics[angle=-90]{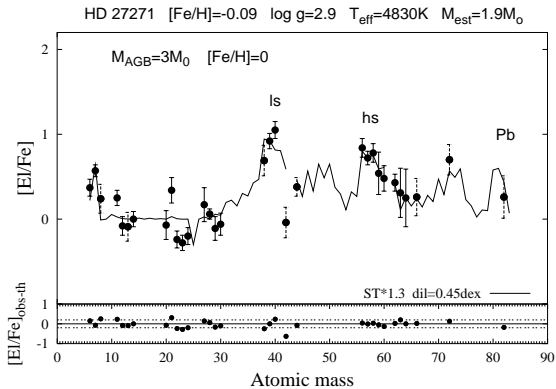}}
\caption{Barium giant HD 27271. Fit obtained with an AGB model of 3 M$_\odot$ initial mass, 
[Fe/H] = 0, ST $\times$ 1.3 $^{13}$C-pocket efficiency and 0.45 dex dilution. }
 \label{hd27271m3}
\end{figure}

In Table~\ref{tabba}, the major characteristics of the 
Allen \& Barbuy (2006) sample are shown.
For each star, we report the effective temperature, 
luminosity, log $g$, the metallicity, the estimated
mass of the star, 
the convective envelope mass according to FRANEC (see also Izzard \& Glebbeek 2006),
and the parameters used to obtain the theoretical 
fits (efficiency of the $^{13}$C-pocket, initial mass of the AGB, 
dilution factor and reduced $\chi$-squared $\bar{\chi^2}$ to evaluate the goodness of the fit).
Note as, in general, the adopted $^{13}$C-pocket does not change with
the initial AGB mass (Col.~9). The only exception is HD 
76225, with a case ST for a model $M$ = 1.5 $M_\odot$ and
ST/1.5 for models $M$ = 2 and 3 $M_\odot$.
For the two giants HD 749 and HD 12392, a dilution of about 0.5 dex
can be obtained by decreasing the metallicity by about 0.2 dex.

\begin{figure}[htbp]
  \resizebox{75mm}{!}{\includegraphics[angle=-90]{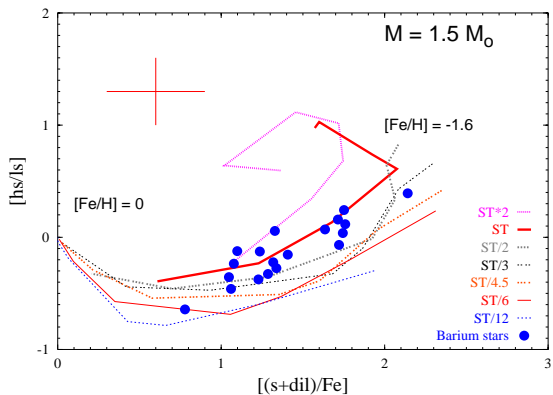}}
\caption{Theoretical results for [hs/ls] versus [(s+dil)/Fe] (where
dil is the dilution factor), using AGB 
models of initial mass $M^{\rm AGB}_{\rm ini}$ = 1.5 $M_{\odot}$
and a range of $^{13}$C-pocket efficiencies. Full circles
correspond to spectroscopic observations of barium stars 
of the Allen \& Barbuy (2006) sample. 
For higher $^{13}$C-pocket efficiencies and decreasing the 
metallicity, [(s+dil)/Fe] decreases because the s-process feeds mostly 
Pb.}
 \label{hustipasa_sara3}
\end{figure}

\begin{figure}[htbp]
  \resizebox{75mm}{!}{\includegraphics[angle=-90]{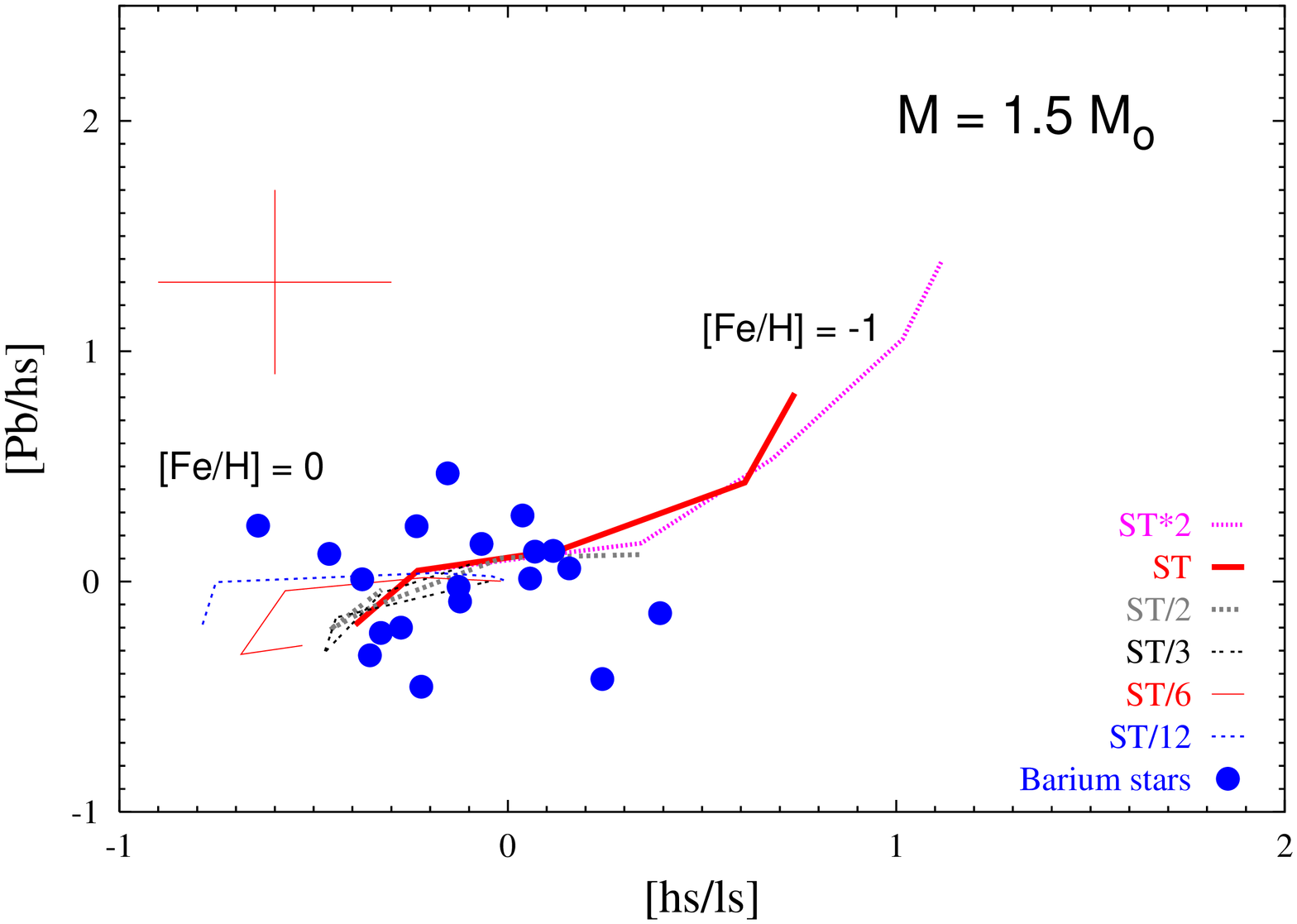}}
\caption{The same as Fig.~\ref{hustipasa_sara3}, but for
[Pb/hs] versus [hs/ls].}
 \label{hustipasa_sara4}
\end{figure}

In Fig.~\ref{hustipasa_sara3} we show theoretical predictions
of [hs/ls] versus [(s+dil)/Fe] for AGB models of $M$ = 1.5 $M\odot$. 
Here '[s/Fe]' means ([ls/Fe]+[hs/Fe])/2 and
'dil' is the dilution factor used in the model. 
Stars interpreted with a $M^{\rm AGB}_{\rm ini}$ = 1.5 $M_\odot$ are 
represented with full circles (see Table~\ref{tabba}, Cols.~10 and~11).
Typical errorbars are shown.
Similarly, in Fig.~\ref{hustipasa_sara4}, we shown theoretical predictions
as compared with observations of [Pb/hs] versus [hs/ls]. 
[Pb/hs] shows quite a large scatter with respect to predictions, 
due to the fact that only one line of Pb has been detected.

For more detailed descriptions on the other stars we refer to
Husti PhD Thesis (2008), and to the preliminary analyses by
Husti \& Gallino (2008) and Husti, Gallino \& Straniero (2008).

Smiljanic et al. (2007) analysed a sample of 8 classical and mild 
barium giants, known in the literature, reanalysed at high-resolution
and S/N. 
The estimated mass is in the range 1.9 to 4.2 
$M_\odot$ and the metallicity ranging from solar down to [Fe/H] $\sim$ $-$0.3.
For most of these giants, only one line was detected for Sr and Sm, and
two lines for Nd. For these elements no errorbars are given by the authors.
Y, Zr, La and Ce are better determined.
The quoted errorbars are typically higher in the Smiljanic 
et al. (2007) sample than in Allen \& Barbuy (2006), 
despite higher S/N spectra.  
No lead abundances were determined.
A preliminary analysis of the two barium stars HD 116713 and HD 205011 
has been presented by Husti \& Gallino (2008).
In Table~\ref{tabsmi}, we show the main characteristics of this sample,
with their theoretical interpretations, in particular, our [Pb/Fe] predictions.
For the two stars HD 46407 and HD 204075, a dilution of 0.7 dex and [Pb/Fe] 
$\sim$ 1.4 would be obtained by decreasing the metallicity by $\sim$ 0.2 dex. 
All stars have a typical [hs/ls] close to 0.
The only exception is HD 139195 for which [hs/ls] $\sim$ 0.25.

\section{Mass transfer calculations}

Boffin \& Jorissen (1988) showed that the amount of mass 
transferred from an AGB star to its companion through stellar 
winds ($\Delta M_2$) depends on the present value of the 
orbital separation, the mass of the barium star and of the 
white dwarf companion:
\begin{eqnarray}
\Delta M_2=\frac{1}{\sqrt{1-e^2}}\left[\frac{GM_2}{V_{wind}^2}\right]^2\frac{\alpha}{2A}\Delta M_1 \nonumber \\
\times\left[\frac{1}{1+G(M_1+M_2)/AV_{wind}^2}\right]^{3/2}, 
\end{eqnarray}
where $M_1$ is the mass of the white dwarf, $M_2$ the mass of 
the barium star, $\Delta M_1$ is the mass lost by winds by the 
AGB companion, $A$ is the semi-major axis, $\alpha$ is a free parameter 
put equal to 2 (Boffin \& Jorissen 1988) and $e$ the eccentricity.
The semi-major axis $A$ varies with the mass exchange between 
the two binaries. A choice $A$ = constant may be considered an 
acceptable approximation during the superwind phase at the end 
of the AGB.
Our FRANEC models adopt the Reimers (1977) mass loss formula 
and do not explicitly include superwinds, however we recall that 
almost 0.5 M$_\odot$ of the envelope mass is ejected with the 
same s-process composition.

\begin{figure}[htbp]
  \resizebox{75mm}{!}{\includegraphics[angle=-90]{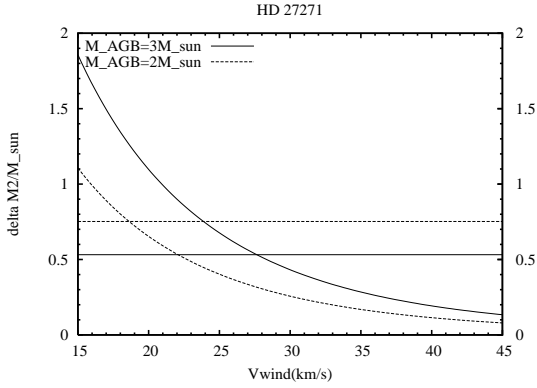}}
\caption{Transferred mass onto the barium giant HD 27271 
versus wind velocity.}
 \label{hd27271mt}
\end{figure}

Thus knowing the mass of the convective envelope from FRANEC
and the dilution factor obtained from the abundance fits we may derive the 
velocity of the wind, for the giants and subgiants, of which
we know the orbital period, by comparing $\Delta M_2$ from the 
equation above with $M^{transf}_{\rm AGB}$ from Section 2.

\begin{figure}[htbp]
  \resizebox{75mm}{!}{\includegraphics[angle=-90]{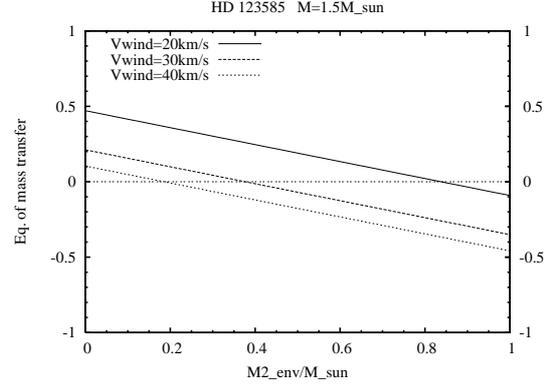}}
\caption{Equation of the mass transfer versus mass of the 
envelope mixed with the s-enhanced material, for star HD 
123585, the case of the 1.5 M$_\odot$ initial AGB mass of 
the companion.}
 \label{hd123585m1p5mt}
\end{figure}

In Fig.~\ref{hd27271mt} we plot $\Delta M_2$ $(\equiv 
M^{transf}_{\rm AGB})$ as a function of 
the wind velocity and the value of $M^{transf}_{\rm AGB}$ 
$(\equiv \Delta M_2)$ obtained from the dilution factor used to 
get the fit. For HD 27271 we obtain 2 solutions: $V_{wind} 
\simeq$ 23 km/s if we consider the initial mass of the AGB
to be 3 M$_\odot$, and  $V_{wind} \simeq$ 18 km/s if we 
consider the initial mass of the AGB to be 2 M$_\odot$. 
The wind velocity for AGB stars was estimated to be in 
the interval 20 $-$ 40 km/s (Knapp 1985; Pottasch 1984). 
Only the solution obtained for the 3 M$_\odot$ case is 
in this range.

\begin{figure}[htbp]
  \resizebox{75mm}{!}{\includegraphics[angle=-90]{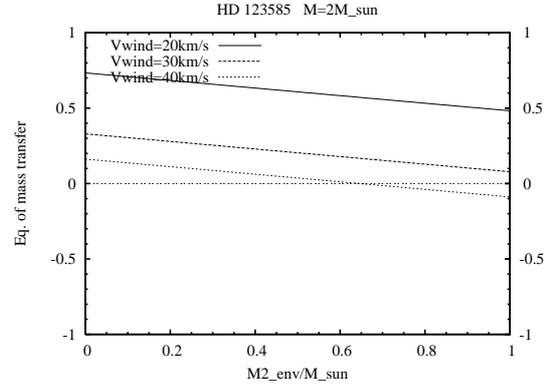}}
\caption{Equation of the mass transfer versus mass of 
the envelope mixed with the s-enhanced material, for star 
HD 123585, the case of the 2 M$_\odot$ initial AGB mass of 
the companion.}
 \label{hd123585m2mt}
\end{figure}

\begin{figure}[htbp]
  \resizebox{75mm}{!}{\includegraphics[angle=-90]{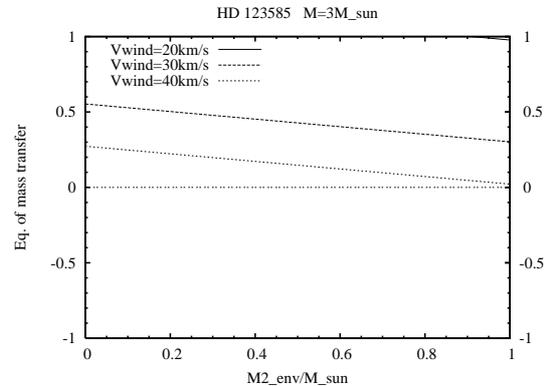}}
\caption{Equation of the mass transfer versus mass of the 
envelope mixed with the s-enhanced material, for star HD 
123585, the case of the 3 M$_\odot$ initial AGB mass of the companion.}
 \label{hd123585m3mt} 
\end{figure}

Half of the sample are barium dwarfs.
The range of estimated mass of dwarfs by Allen \& 
Barbuy (2006) is between 0.9 $\leq$ $M_{est}$/$M_\odot$ 
$\leq$ 1.4.
No convective envelope is expected for these objects.
Many of the stars need dilution factor higher than 0.5 dex.
This might be explained by moderate to strong thermohaline 
mixing.
Putting to zero the mass transfer equation, and replacing 
$\Delta M_2$ with $M^{env}_{\star}/10^{dil}$, we may 
estimate the depth of the mixing. In Fig.s~\ref{hd123585m1p5mt},
~\ref{hd123585m2mt} and~\ref{hd123585m3mt},
we have plotted the equation of mass transfer versus the 
amount of envelope mass over which the s-rich material is 
diluted due to thermohaline mixing, for HD 123585. In the 
2 and 3 M$_\odot$ cases, the depth of the mixing (in mass) 
would exceed the mass of the barium star for $V_{wind}$ = 
20 km/s, so we discard these cases. For the 1.5  M$_\odot$, 
the amount of envelope mass mixed with the s-rich transferred 
material is in the range of $\sim$ 0.20 and $\sim$ 0.85 
M$_\odot$.

In Table \ref{gsg} we report the period and the eccentricity available in the literature for 
six of the barium stars of our sample (Udry et al. 1998a,b $-$ HD 5424, HD 22589, HD 27271 and Pourbaix \& Jorissen 2000 $-$ HD 89948, HD 107574, HD 123585), as well as the calculated wind velocity for giants (or subgiants), having various masses of the AGB donor. For HD 5424 we obtain two solutions in the range of 20-40 km/s, for HD 25589 and HD 27271 we have only one solution in the given range, while for HD 107574 we have no solution in the chosen interval, the closest one is for a 1.5 M$_\odot$ AGB companion, with V$_{wind}$ = 42 km/s.

In Tables \ref{hd89948} and \ref{hd123585} we present the depth in mass of the  thermohaline mixing for the two dwarf stars of the sample with known orbital period. The depth of the mixing depends on the mass of the AGB and on the assumed wind velocity. For HD 89948 we obtain similar results as for HD 123585. The mixing seems to be deeper for the lower metallicity star. 

\begin{sidewaystable*}[]                                                                                                                                                                                                                                                  
\caption{Stellar parameters and results for the Allen \& Barbuy (2006) sample.
$\bar{\chi^2}$ is calculated for the s-process elements Y, Zr, La, Nd, Sm 
and Pb. The highest values are principally due to the lead uncertainty.}                                                                                                                                                                                                                                              
\begin{tabular}{|l|c|c|c|c|c|c|c|c|c|c|c|c|c|c|c|c|}                                                                                                                                                                                                                          
\hline                                                                                                                                                                                                                                                                    
Stars	  & $T_{\rm eff}$ &	log $T_{\rm eff}$	 	& log ($L/L_\odot$)	& log $g$	& [Fe/H]&  $M^{\rm est}$/$M_\odot$	&  $M^{env}_{obs}$/$M_\odot$  &  CASE & \multicolumn{2}{c|}{$M$=1.5$M_\odot$} & \multicolumn{2}{c|}{$M$=2.0$M_\odot$} & \multicolumn{2}{c|}{$M$=3.0$M_\odot$} \\
\cline{10-15}                                                                                                                                                                                                                                                             
        &      &                     &             &       &       &         &     &      & dil & $\bar{\chi^2}$ & dil & $\bar{\chi^2}$ &  dil & $\bar{\chi^2}$ \\                                                                                                                                       
\hline                                                                                                                                                                                                                                                                    
HD 749	  & 4610 &	3.663	&  1.322     	& 2.8	& -0.06	&  1.2	    &  0.9  &     ST$\times$1.3  & -     &  -        & -   &  -     &      0  & 2.7 \\                                                                                                                                          
HR 107	  & 6440 &	3.808	&  0.623     	& 4.08	& -0.36	&  1.2	    &  0    &     ST$\times$1.3 & 0.9   &  1.5      & 1.2 &  1.3   &      1.3    & 1.1 \\                                                                                                          
HD 5424	  & 4570 &	3.659	&  2.264     	& 2		& -0.55	&  1.9	    &  1.6  &     ST            & -     &  -        & 0.8 &  3.7   &      0.7    & 3.3\\                                                                                                                                                       
HD 8270	  & 5940 &	3.773	&  0.204     	& 4.2	& -0.42	&  0.9	    &  0    &     ST/2          & 0.5   &  1.0      & 0.8 &  1.0   &      0.7    & 1.6 \\                                                                                                                                             
HD 12392	  & 5000 &	3.698	&  1.301     	& 3.2	& -0.12	&  2	    &  1.5  &      ST$\times$2    & -     &  -        & -   &  -     &      0  & 5.6 \\                                                                                                                                          
HD 13551	  & 5870 &	3.768	&  0.361     	& 4		& -0.44	&  0.9	    &  0    &     ST/2          & 0.4   &  1.0      & 0.7 &  1.5   &      0.7    & 1.2 \\                                                                                                                                             
HD 22589	  & 5400 &	3.732	&  1.204     	& 3.4	& -0.27	&  1.6	    &  0.5  &     ST/2             & -     & -         & 0.4 &  2.9   &      0.5    & 2.5 \\                                                                                                                                                  
HD 27271	  & 4830 &	3.683	&  1.505     	& 2.9	& -0.09	&  1.9	    &  1.5  &    ST$\times$1.3       & -     & -         & 0.3 &  1.5   &      0.45   & 1.5 \\                                                                                                                            
HD 48565	  & 5860 &	3.767	&  0.397     	& 4.01	& -0.62	&  0.9	    &  0    &     ST            & 0.6   &  0.4      & 0.9 &  0.5   &      0.8    & 0.5 \\                                                                                                                                                  
HD 76225	  & 6110 &	3.786	&  0.903     	& 3.8	& -0.31	&  1.4	    &  0    &     ST--ST/1.5        & 0.2   &  3.6      & 0.2 &  1.6   &      0.2    & 2.6 \\                                                                                                                                          
HD 87080	  & 5460 &	3.737	&  0.698     	& 3.7	& -0.44	&  1.2	    &  0.1  &      ST           & 0.4   &  2.5      & 0.7 &  2.6   &      0.7    & 2.0 \\                                                                                                                                                
HD 89948	  & 6010 &	3.778	&  0.204     	& 4.3	& -0.3	&  1	    &  0    &     ST/1.5        & 0.2   &  1.7      & 0.5 &  0.8   &      0.5    & 0.8 \\                                                                                                                                         
HD 92545	  & 6210 &	3.793	&  0.698     	& 4		& -0.12	&  1.3	    &  0    &     ST$\times$2   & 0.5   &  1.7      & 0.7 &  1.3   &      0.8    & 1.9 \\                                                                                                                  
HD 106191  &	5890 &	3.770	&  0.278     	& 4.2	& -0.29	&  1	    &  0    &     ST/1.5        & 0.3   &  2.2      & 0.6 &  2.2   &      0.6    & 3.2 \\                                                                                                                                        
HD 107574  &	6400 &	3.806	&  1.113     	& 3.6	& -0.55	&  1.4	    &  0.3  &     ST            & 0.8   &  2.7      & 1.1 &  3.0   &      1.0    & 2.8 \\                                                                                                                                                 
HD 116869  &	4720 &	3.673	&  1.944     	& 2.2	& -0.32	&  1.2	    &  0.9  &      ST$\times$2  & 1.0   &  0.6      & 1.2 &  0.6   &      1.3    & 0.6 \\                                                                                                                 
HD 123396  &	4360 &	3.639	&  2.406     	& 1.4	& -1.19	&  0.8	    &  0.45 &     ST/4.5        & 1.0   &  3.9      & 1.2 &  4.0   &      -      & - \\                                                                                                                                             
HD 123585  &	6350 &	3.802	&  0.447     	& 4.2	& -0.48	&  1.1	    &  0    &     ST            & 0.25  &  0.7      & 0.6 &  0.3   &      0.5    & 0.7 \\                                                                                                                                                
HD 147609  &	5960 &	3.775	&  0.079     	& 4.42	& -0.45	&  1	    &  0    &     ST/2             &  -    & -         & 0.2 &  1.6   &      0.1    & 2.1 \\                                                                                                                                                     
HD 150862  &	6310 &	3.800	&  0.041     	& 4.6	& -0.10	&  1.1	    &  0    &     ST$\times$1.3 & 0     &  8.8      & 0.2 &  10.0  &      0.3    & 9.0 \\                                                                                                             
HD 188985  &	6090 &	3.784	&  0.278     	& 4.3	& -0.3	&  1.1	    &  0    &     ST            & 0.2   &  1.2      & 0.5 &  1.1   &      0.6    & 1.7 \\                                                                                                                                                   
HD 210709  &	4630 &	3.665	&  1.681     	& 2.4	& -0.04	&  1.1	    &  0.9  &     ST$\times$2   & 0.5   &  0.8      & 0.7 &  0.6   &      0.9    & 0.5 \\                                                                                                                    
HD 210910  &	4570 &	3.659	&  1.278     	& 2.7	& -0.37	&  1	    &  0.7  &     ST/1.5        & 0.4   &  4.7      & 0.7 &  4.4   &      0.8    & 4.1 \\                                                                                                                                         
HD 222349  &	6130 &	3.787	&  0.698     	& 3.9	& -0.63	&  1.2	    &  0    &     ST            & 0.6   &  1.1      & 0.9 &  1.1   &      0.8    & 1.2 \\                                                                                                                                                  
BD +185215 &	6300 &	3.799	&  0.477     	& 4.2	& -0.53	&  1.1	    &  0    &     ST/2          & 0.3   &  4.6      & 0.6 &  4.5   &      0.6    & 2.9 \\                                                                                                                                            
HD 223938  &	4970 &	3.696	&  1.255     	& 3.1	& -0.35	&  1.4	    &  1.1  &     ST            & 0.4   &  3.6      & 0.7 &  3.2   &      0.8    & 3.1 \\                                                                                                                                                   
\hline                                                                                                                                                                                                                                                       
\end{tabular}                                                                                                                                                                                                                                                             
\label{tabba}                                                                                                                                                                                                                                                             
\end{sidewaystable*}                                                                                                                                                                                                                                                      

\small{
\begin{table*}[]
\begin{center}
\caption{[ls/Fe], [hs/Fe], [Pb/Fe] and their ratios [hs/ls] and [Pb/hs]
for the Allen \& Barbuy (2006) sample.}
\label{hsls}
\begin{tabular}{|l|c|c|c|c|c|c|}
\hline 
Stars    &   [Fe/H]	 &   [ls/Fe]&	[hs/Fe]	 &  [Pb/Fe]	 &   [hs/ls]&	[Pb/hs] \\
\hline
HD 749	    &   -0.06	 &   1.32	&  1.17	     &   0.38	 &    -0.14	&   -0.79   \\ 
HR 107	    &   -0.36	 &   0.59	&  0.43	     &   0.90	 &    -0.16	&   0.47    \\ 
HD 5424	    &   -0.55	 &   1.19	&  1.51	     &   1.10	 &    0.32	&   -0.41   \\ 
HD 8270	    &   -0.42	 &   0.98	&  0.70	     &   0.50	 &    -0.28	&   -0.20   \\ 
HD 12392	    &   -0.12	 &   1.29	&  1.52	     &   1.15	 &    0.24	&   -0.37   \\ 
HD 13551	    &   -0.44	 &   1.05	&  0.72	     &   0.50	 &    -0.33	&   -0.22   \\ 
HD 22589	    &   -0.27	 &   0.95	&  0.36	     &   -0.15	 &    -0.59	&   -0.51   \\ 
HD 27271	    &   -0.09	 &   0.97	&  0.55	     &   0.31	 &    -0.41	&   -0.24   \\ 
HD 48565	    &   -0.62	 &   1.10	&  1.22	     &   1.35	 &    0.12	&   0.13    \\ 
HD 76225	    &   -0.31	 &   1.22	&  0.84	     &   0.85	 &    -0.38	&   0.01    \\ 
HD 87080	    &   -0.44	 &   1.23	&  1.47	     &   1.05	 &    0.24	&   -0.42   \\ 
HD 89948	    &   -0.30	 &   1.03	&  0.67	     &   0.35	 &    -0.36	&   -0.32   \\ 
HD 92545	    &   -0.12	 &   0.70	&  0.46	     &   0.70	 &    -0.24	&   0.24    \\ 
HD 106191	&   -0.29	 &   0.99	&  0.53	     &   0.65	 &    -0.46	&   0.12    \\ 
HD 107574	&   -0.55	 &   0.96	&  0.89	     &   1.05	 &    -0.07	&   0.16    \\ 
HD 116869	&   -0.32	 &   0.64	&  0.79	     &   0.85	 &    0.16	&   0.06    \\ 
HD 123396	&   -1.19	 &   0.95	&  1.34	     &   1.20	 &    0.39	&   -0.14   \\ 
HD 123585	&   -0.48	 &   1.35	&  1.42	     &   1.55	 &    0.07	&   0.13    \\ 
HD 147609	&   -0.45	 &   1.57	&  1.35	     &   0.78	 &    -0.22	&   -0.57   \\ 
HD 150862	&   -0.10	 &   1.10	&  0.46	     &   0.70	 &    -0.64	&   0.24    \\ 
HD 188985	&   -0.30	 &   1.10	&  0.97	     &   0.95	 &    -0.13	&   -0.02   \\ 
HD 210709	&   -0.04	 &   0.66	&  0.54	     &   0.45	 &    -0.12	&   -0.09   \\ 
HD 210910	&   -0.37	 &   0.45	&  0.57	     &   -0.04	 &    0.12	&   -0.61   \\ 
HD 222349	&   -0.63	 &   1.13	&  1.16	     &   1.45	 &    0.04	&   0.29    \\ 
BD +185215	&   -0.53	 &   1.13	&  0.91	     &   0.45	 &    -0.22	&   -0.46   \\
HD 223938	&   -0.35	 &   0.90	&  0.96	     &   0.97	 &    0.06	&   0.01    \\ 
\hline
\end{tabular}
\end{center}
\end{table*}
}

\small{
\begin{table*}[]
\begin{center}
\caption{Stellar parameters and results for the Smiljanic et al. (2007) sample.}
\label{tabsmi}
\begin{tabular}{|l|c|c|c|c|c|c|c|c|}
\hline
Stars  &    [Fe/H]  &$T_{\rm eff}$ 	&  log $g$  &  $M_{\rm est}$  &	CASE				& $M$=3$M_\odot$ &    $M$=4$M_\odot$ &	[Pb/Fe]$_{\rm th}$ \\
 		   &  			&			&		      &                &                       &   dil &   dil &              \\
\hline  
                                                            
HD 46407  	&  	-0.09	&  4940		&    2.65	  &     2.3           &  ST$\times$2	   &	0.2&   -  &     1.1        \\
HD 104979 	&  	-0.35	&  4920		&    2.58	  &     2.3           &	ST$\times$1.3	   &	1.4	        &   -  &     0.6            \\
HD 116713   &    -0.12	&  4790		&    2.67	  &     1.9           &	ST$\times$2		   &   0.5		    &   -  &     0.8	           \\
HD 139195   &    -0.04	&  4910		&    2.41	  &     2.4           &	ST           	   &	0.8 		&   -  &     0.0            \\
HD 181053   &    -0.35	&  4920		&    2.58	  &     2.2           &	ST				   &	1.7		    &   -  &     0.2            \\
HD 202109   &    -0.04	&  4910		&    2.41	  &     3.0           &	ST$\times$2		   &	1.5		    &   -  &     0.2	           \\
HD 204075   &    -0.09	&  5250		&    2.49	  &     4.2           &	ST$\times$2   &   -            & 0.3 &     1.2        \\
HD 205011   &    -0.14	&  4780		&    2.41	  &     2.2           &	ST$\times$1.3	   &	0.6		    &   -  &     0.4            \\
\hline
\end{tabular}
\end{center}
\end{table*}
}

\small{
\begin{table*}[]
\begin{center}
\caption{Parameters of the theoretical solutions. Calculated wind velocities for giants and subgiants.}
\label{gsg}
\begin{tabular}{|l|c|c|c|c|c|}
\hline Stars & P (days) & e &$M_{\rm AGB}$=3$M_\odot$ & $M_{\rm AGB}=$2$M_\odot$ & $M_{\rm AGB}$=1.5$M_\odot$   \\
\cline{4-6}
 & & &$V_{\rm wind}$&$V_{\rm wind}$&$V_{\rm wind}$\\
 & & & (km/s) & (km/s) &  (km/s)\\
\hline
HD 5424   & 1881 & 0.225 & 32 & - & - \\
HD 22589  & 5721 & 0.240 & 22 &  17 & - \\
HD 27271  & 1694 & 0.217 & 23 &  18 & - \\
HD 89948 &  668 & 0.117 & - & - & - \\
HD 107574  & 1350 & 0.081 & $>$45 & $>$45 & 42 \\
HD 123585 & 458 & 0.062 & - & - & - \\
\hline
\end{tabular}
\end{center}
\end{table*}
}

\begin{table*}[]
\begin{center}
\caption{HD 89948. Depth of the thermohaline mixing.}
\label{hd89948}
\begin{tabular}{|l|c|c|c|}
\hline HD 89948 & $M_{\rm AGB}$ = 1.5$M_\odot$ & $M_{\rm AGB}$ = 2$M_\odot$ & $M_{\rm AGB}$ = 3$M_\odot$    \\
\hline
$V_{\rm wind}$ = 20 km/s& 0.5& $>$1 & $>$1   \\
$V_{\rm wind}$ = 30 km/s& 0.22& 0.7 & $>$1   \\
$V_{\rm wind}$ = 40 km/s& 0.10& 0.3 & 0.55   \\
\hline
\end{tabular}
\end{center}
\end{table*}

\begin{table*}[]
\begin{center}
\caption{HD 123585. Depth of the thermohaline mixing.}
\label{hd123585}
\begin{tabular}{|l|c|c|c|}
\hline HD 123585 & $M_{\rm AGB}$ = 1.5$M_\odot$ & $M_{\rm AGB}$ = 2$M_\odot$ & $M_{\rm AGB}$ = 3$M_\odot$   \\
\hline
$V_{\rm wind}$ = 20 km/s & 0.83 & $>$1 &$>$1  \\
$V_{\rm wind}$ = 30 km/s & 0.37 & $>$1 & $>$1 \\
$V_{\rm wind}$ = 40 km/s & 0.18 & 0.65 &$>$1  \\
\hline
\end{tabular}
\end{center}
\end{table*}

\section{Conclusions}
We have found theoretical interpretations for the abundance 
patterns of 26 barium dwarfs, barium subgiants and giants 
from Allen \& Barbuy (2006) and of 8 barium giants from 
Smiljanic et al. (2007). In order to obtain these 
solutions, we have applied various dilution factors 
(from 0 to 1.5 dex) due to the mixing of the s-enhanced 
material with the envelope of the accretor. In most cases, 
we have not found unique solutions: models 
having the same metallicity and $^{13}$C-pocket efficiency, 
but different masses and dilution factors match just as 
well the observations.

For barium giants and subgiants that have large convective 
envelopes, an appreciable dilution factor has to be expected,
as confirmed by our results. 
For several barium dwarfs that do not have a convective 
envelope, an important dilution factor must also be applied. 
This comes as a proof that efficient thermohaline mixing is active 
in these stars (Stancliffe et al. 2007). 

We have performed some mass transfer calculations for 
the stars that have the orbital period reported in 
literature. For barium giants and subgiants we could thus obtain 
the stellar wind velocity, while for barium dwarfs, assuming a 
certain initial mass of the AGB and a certain value of the
wind velocity, we could estimate the depth of the 
thermohaline mixing. 

Further observations of barium stars with high-resolution
spectra and S/N would be very helpful.

\section*{Acknowledgments} 
We thank the referee Rob Izzard for his incisive suggestions
to improve the paper.
Work supported by the Italian MIUR$-$PRIN 2006 Project
`Late Phases of Stellar Evolution: Nucleosynthesis in Supernovae,
AGB Stars, Planetary Nebulae'.


\end{document}